\documentclass[12pt]{iopart}
\usepackage{amssymb,graphicx}

\newcommand{\be}{\begin{equation}}
\newcommand{\ee}{\end{equation}}
\newcommand{\bea}{\begin{eqnarray}}
\newcommand{\eea}{\end{eqnarray}}
\renewcommand{\(}{\left (}
\renewcommand{\)}{\right )}

\renewcommand{\Re}{\mathop{\rm Re}}
\renewcommand{\Im}{\mathop{\rm Im}}
\newcommand\mpl{m_{\rm pl}}
\newcommand{\vp}{\varphi}
\newcommand{\Lkin}{\mathcal{L}_\mathrm{kin}}
\newcommand{\GeV}{\, \mathrm{GeV}}

\begin{document}

\hfill DESY 08-004

\title{SUGRA chaotic inflation and moduli stabilisation}

\author{S~C~Davis$^1$  and M~Postma$^{2,3}$}

\address{${}^1$ Service de Physique Th\'eorique, 
Orme des Merisiers, CEA/Saclay, 91191 Gif-sur-Yvette Cedex, France}
\address{${}^2$ DESY, Notkestra\ss e 85, 22607 Hamburg, Germany}
\address{${}^3$ Nikhef, Kruislaan 409, 1098 SJ Amsterdam, The
Netherlands}

\eads{\mailto{sdavis@lorentz.leidenuniv.nl},
\mailto{postma@mail.desy.de}}

\begin{abstract}
Chaotic inflation predicts a large gravitational wave signal which can be
tested by the upcoming Planck satellite.  We discuss a SUGRA implementation of
chaotic inflation in the presence of moduli fields, and find that
inflation does not work with a generic KKLT moduli stabilisation
potential. A viable model can be constructed with a fine-tuned 
moduli sector, but only for a very specific choice of K\"ahler potential.
Our 
analysis also shows that inflation models satisfying 
$\partial_{i} W_{\rm inf}=0$ for all inflation sector fields $\phi_i$
can be combined successfully with a fine-tuned moduli sector.
\end{abstract}

\section{Introduction}

In chaotic inflation models the energy scale of inflation is high, typically
of the order of the grand unified scale~\cite{linde}. As a consequence  these
models give a large tensor contribution to the density perturbations.  This
makes them testable by current and future CMB experiments, most notably by the
upcoming Planck satellite.  However, chaotic inflation is not easy to
implement in a supergravity theory~\cite{kawasaki,kadota}.  The inclusion of other
high energy physics, such as moduli fields, creates further
problems~\cite{brax,kallosh,kallosh2,kss}. Naturally, any realistic inflation
model must be part of some full theory, containing all known physics. The
effects of other sectors of the theory on inflation can not be
ignored.

As shown by Lyth~\cite{lyth} in the context of slow-roll inflation, a
measurable tensor mode requires the inflaton field to change by superplanckian
values during inflation. 
Examples of such ``large field
models'' of inflation  are chaotic and natural inflation~\cite{natural}.  At
present no string theory derivation of a large field inflaton model exists.
The displacement of the inflaton in brane models of inflation is bounded by
the size of the compactified space, and in all known models less than the
Planck scale~\cite{baumann,bean}.  In all examples of modular inflation the
inflationary scale is too low for an appreciable tensor
signal~\cite{kallosh,kallosh2}. N-flation~\cite{kss,Nflation,grimm}, the stringy realization of assisted
inflation~\cite{assisted}, 
gives rise to appreciable tensor modes.   However, it is not clear whether all
of the underlying assumptions are satisfied in these models \cite{kallosh}.
Despite the negative results, so far there is no  ``no-go'' theorem
stating that string theory cannot give large field inflation.  It may very
well be that it can be realized in corners of the landscape not yet explored
-- after all, the search has only just begun.

In this paper we consider a ${\mathcal N} = 1$ SUGRA implementation of chaotic
inflation, and analyse what happens when it is combined with a KKLT-like
moduli sector.  In our set-up the inflaton and moduli sector only interact
gravitationally.   Our approach is phenomenological in that we analyse the
SUGRA effective field theory, but do not attempt to derive the model from
string theory.    It should be noted in this context that moduli fields are
not unique to string theory.  Flat directions abound in any SUSY theory.  If
SUSY is broken in some hidden sector by non-perturbative physics,  the moduli
sector has the same qualitative properties as the KKLT model, and our results
 apply.  

As mentioned,  it is not easy to construct a model of chaotic inflation  in the
presence of additional moduli fields, even when they are stable.  First of
all, there is the $\eta$-problem, present in all models of SUGRA
inflation~\cite{copeland,dine}.  The potential during inflation is of the form
$V \sim \e^K \tilde V$, which for a canonically normalised inflaton field
$\vp$ gives rise to a large inflaton mass $m_\vp^2 \sim H^2$ ruining slow-roll
inflation.  This can be solved by fine-tuning the K\"ahler potential so that
the inflaton mass is accidentally small.  More elegantly, the inflaton mass
can be protected by symmetries.  In this paper we will introduce a shift
symmetry for the inflaton field that leaves the K\"ahler potential invariant
to solve the $\eta$-problem~\cite{kadota,gaillard,banks}.

Inclusion of moduli fields in the system gives rise to a whole new set
of obstacles to implement inflation.  The moduli fixing potential
breaks supersymmetry.  Consequently there are soft corrections to the
inflaton potential.  The soft terms are small in the limit of low
scale SUSY breaking, with a small gravitino mass $m_{3/2}^2 \lesssim
H^2$.  At the same time, the requirement that the moduli fields
remain stabilised in their minimum during inflation, and do not run away to
infinity, implies that the moduli masses should be sufficiently large.
This requirement is usually expressed as a constraint on the Hubble
parameter during inflation $H^2 \ll m_{\rm mod}^2$~\cite{KL}.  In a generic
potential $m_{\rm mod}^2 \sim m_{3/2}^2$ and without fine-tuning
(in addition to that required to set the cosmological constant
to zero) these requirements are at odds with each other.  This is for
example the case in the original KKLT model~\cite{KKLT}.  It is difficult to
embed large field inflation in such a set up.

We see there is tension between keeping the soft corrections to the inflaton
potential small, and keeping the moduli fields fixed during inflation. This
can be eased if the modulus sector is fine-tuned so that the modulus and
gravitino masses are no longer of the same order of magnitude.  This is
achieved explicitly in the Kallosh-Linde (KL) set-up~\cite{KL}, which uses a
racetrack potential for the modulus field.  In this case, parameters are tuned
so that the modulus mass is much larger than the gravitino mass.  Having the
Hubble constant during inflation  between these mass scales 
$m_{\rm mod}^2 \ll H^2 \ll m_{3/2}^2$ offers a way to solve both problems.
Note that it also allows the gravitino mass to be in the phenomenologically
favoured TeV range, without the need for low scale inflation (in fact, this was
the original motivation for KL).

In this paper we will analyse chaotic inflation in the presence of a
single modulus field with a no-scale K\"ahler potential. 
The models we will study have the superpotential  
\be
W = W_{\rm mod}(T) + m \phi_1 \phi_2 \, .
\label{w1}
\ee
We consider both a generic KKLT potential and a fine-tuned KL potential.   The
above inflaton superpotential was first  proposed in~\cite{kawasaki}.
Refs.~\cite{brax,kallosh,kallosh2} added a moduli sector to the set-up.  We
extend their results by an in-depth discussion of the effects of the moduli
dynamics, with an emphasis on finding the conditions for successful inflation.
As expected, inflation does not work in the KKLT set-up.  Whether KL works
depends sensitively on the K\"ahler potential for the inflaton
fields. Although the moduli corrections are small after inflation due to the
fine-tuning in the KL set-up, this is not necessarily true during inflation.
During inflation the modulus field $T$ is slightly displaced from its
post-inflationary minimum, disrupting the minute fine-tuning of the potential,
with potentially large effects.   Indeed,  consider the following K\"ahler
potentials
\numparts
\bea
K_{1} &=& -3\log[T+\bar T] -\frac12 (\phi_1 -\bar{\phi}_1)^2 
+ \phi_2 \bar{\phi}_2 \, ,
\label{k1}  \\
K_{2} &=& -3\log[T+\bar T] -\frac12 (\phi_1 -\bar{\phi}_1)^2 
-\frac12( \phi_2- \bar{\phi}_2)^2 \, ,
 \label{k2}  \\
K_{\alpha} &=& -3\log\left[T+\bar T- \frac13 (T+\bar T)^\alpha 
\, \phi_2 \bar{\phi}_2\right]-\frac12 (\phi_1 -\bar{\phi}_1)^2 \, .
\label{kalpha}
\eea
\endnumparts
All K\"ahler potentials have a shift symmetry for the inflaton field
$\phi_1$ to solve the $\eta$-problem.  However, as we will show, only $K_{1}$
combined with the KL modulus sector gives a viable model.  For all the other
models, independent of modular weight $\alpha$, the coupling between the
modulus and inflaton sectors leads to instabilities in the potential, with a
runaway behaviour for some of the fields. It is thus crucial to take the
dynamics of the modulus field during inflation into account for a correct
analysis of the model.

This paper is organised as follows.  The next section provides the
background material, with a concise summary of the KKLT and KL moduli
stabilisation potential, as well as a discussion of SUGRA chaotic
inflation without moduli.  The rest of the paper discusses the
combination of chaotic inflation and moduli fields.  In
section~\ref{s:model2} we study the model with
$K_{2}$~\eref{k2}. Although inflation does not work, it is useful to
analyse why.  In section~\ref{s:model1} we consider the model with
$K_{1}$~\eref{k1}.  As mentioned above, this is a viable model of
chaotic inflation.  We discuss the inflationary predictions, in
particular whether the supergravity corrections can leave a signature
in the CMB.  Finally, in section~\ref{s:KL} we use the insight gained
in the previous sections to discuss more generic combinations  of
chaotic inflation and KL moduli stabilisation, including models
with~\eref{kalpha}.  We end with some concluding remarks.

Throughout this article we will work in units with the reduced Planck mass 
$\mpl = 1/\sqrt{8\pi G_N}$ set to unity.

\section{Background}
\label{s:background}

\subsection{Moduli stabilisation}

Consider a single volume modulus with a no-scale K\"ahler potential $K =
-3\log[T+\bar T]$.  The modulus field is stabilised in an AdS minimum
by a combination of fluxes \cite{gkp} and non-perturbative physics; an
uplifting term is added to end up with a Minkowski vacuum.

\subsubsection{KKLT}

In the original KKLT set-up the superpotential is of the form \cite{KKLT}
\be W= W_0 + A \e^{-a T}.
\label{WKKLT}
\ee
The first term comes from integrating out the complex structure
moduli, the second originates from non-perturbative effects.  The
potential $V_F = \e^K [ W_I K^{I \bar J} W_{\bar J} - 3 |W|^2]$ has a
SUSY AdS minimum.

The lifting term is of the form 
\be
(V_{\rm lift})_1 = \frac{E}{(T+ \bar T)^n}, \quad {\rm or} \quad
(V_{\rm lift})_2 = E \e^K,
\label{Vlift}
\ee
with $E$ a constant which can be tuned to get a zero cosmological
constant.  $(V_{\rm lift})_1$ applies to $D$-term lifting ($n=3$)
\cite{bkq,ana}, or lifting by supersymmetry breaking anti- D3-branes
located in the throat ($n=2$ ) or bulk ($n=3$) \cite{KKLT}. Using
$F$-terms to uplift, the effective potential is of the form 
$(V_{\rm lift})_2$.  An effective $F$-lifting term, as opposed to properly
adding a contribution to $W$ and calculating things through, is only a
good approximation if the SUSY breaking sector is a small correction
to the potential and decouples \cite{KL,GR}. This is not the case for
KKLT, but can be done in a KL set-up discussed below.  This does not
mean that the KKLT potential cannot be uplifted using $F$-terms, but
just that it cannot be described in such a simple way as in
\eref{Vlift}.  The details of the uplifting term do not really matter
for inflation; we checked that using different lifting terms only give
quantitative differences, and in particular it cannot save a sick
model or destroy a healthy one. For definiteness we take 
$(V_{\rm lift})_1$ with $n=2$ in the following.

Without loss of generality we can take $W_0$ to be real and negative.
The potential is then minimised for $\gamma =0$, where we have decomposed the
field into its real and imaginary parts $T = \sigma + \rmi \gamma$. The mass
matrix is
\be
m^i_j = 
g^{ik} (\partial_k \partial_j V - \Gamma^l_{kj} \partial_l V)
+ \frac{2\sigma}{3} (\partial_\sigma + \partial_\gamma)V
\label{massmatrix}
\ee
with $g_{ij}$ the metric on field space spanned by real fields defined
by ${\mathcal L}_{\rm kin} = (1/2) g_{ij}\partial_\mu \phi^i
\partial^\mu \phi^j$.

The derivation of the non-perturbative terms
in~\eref{WKKLT} is only valid for $a \sigma_0 \gg 1$.
In this limit, approximate analytic expressions can be found for
the mass scales~\cite{mfi}. 
The SUSY AdS solution before lifting has $D_T W =0$,
which implies $W_T \approx 3W/(2T)$. This relation survives the
lifting procedure. It then  follows that 
$V_{\rm lift} \approx 3 m_{3/2}^2$, where $m_{3/2}
= \e^{K/2} |W|$ is the gravitino mass. The height of the barrier
preventing the modulus from rolling to infinity is also $3 m_{3/2}^2$, and is
located close to $\sigma= \sigma_0 + (1/a) \log (2 a \sigma_0/n)$. 
Furthermore $W \approx W_0$.
The moduli masses $m_\sigma^2 \approx m_\gamma^2 \approx K^{T \bar T}
V_{T \bar T} \approx (2 a \sigma_0 m_{3/2})^2$ are
somewhat larger than the gravitino mass. For future reference we note
that $W_{TT} \approx -3m_\sigma/\sqrt{2\sigma_0}$.

\subsubsection{KL}

Ref. \cite{KL} constructed a fine-tuned modulus potential with
$m_{3/2} \ll m_\sigma$.  Then for $m_{3/2} \ll H \ll
m_\sigma$ one expects the soft corrections to the inflaton potential
to be small, while the modulus field remains fixed during
inflation. This set-up allows for low-scale SUSY breaking without the
necessity for low-scale inflation.

The idea is to construct a potential that has a supersymmetric
Minkowski vacuum with $W = W_T =0$. Perturbing this potential
slightly, by order $m_{3/2}^2$, gives an AdS minimum with a small
negative cosmological constant.  After uplifting, the result is a small gravitino mass but a large
barrier separating the minimum from the runaway minimum at infinity
(which requires a large modulus mass). 
Lifting can be $F$-term, e.g.\ by
introducing an O'Raifeartaigh sector~\cite{O'Raifeartaigh} as in
\cite{KL}, or by SUSY breaking terms using a throat
$\overline{D3}$-brane.  Implementing a KL-style set-up with $D$-term
lifting does not seem possible, as is produces a barrier height of
similar size to $m_{3/2}^2$.
 
The simplest potential that
does the trick is the modified racetrack potential
\be
W_{\rm mod}= W_0 + A \e^{-a T} - B \e^{-bT} \, .
\label{WKL}
\ee
This has a SUSY Minkowski minimum with $W=W_T=0$ for fine-tuned
parameters
\be W_0 = w_0 \equiv - A \( \frac{bB}{aA} \)^{a/(a-b)} + B \(
\frac{bB}{aA} \)^{b/(a-b)}  \, ,
\label{w_0}
\ee
and with the modulus stabilised at 
\be
\sigma_0 = \frac1{a-b} \ln \( \frac{aA}{bB} \) \, .
\label{sigma0}
\ee
In the limit $a \sigma_0, b \sigma_0 \gg 1$, the maximum of the
potential is located near to $\sigma=\sigma_0 + \ln(a/b)/(a-b)$. For 
$(a-b) \ll a$, its height is then approximately 
$W_0^2 a^2/(6 \sigma_0 \e^2)$. This is of order $m_T^2/(a\sigma_0)^2$,
a relation which incidentally also holds for the KKLT set-up discussed before.

To introduce a non-zero gravitino mass we
perturb $W_0 =w_0 + \delta$ with $\delta \ll 1$. Then $E
\sim \delta^2, W \sim W_T T \sim \delta$.  And thus the
gravitino mass $m_{3/2} \sim \e^{K/2} \delta$ can be made
arbitrarily small. On the other hand  $V_{T \bar T} \sim \e^K
W_0^2$ and thus the moduli mass is insensitive to $\delta$.  The
required hierarchy $m_{3/2}^2 \ll m_\sigma^2$ is obtained when
$\delta^2 \ll K^{T\bar T} W_0^2$.  For future reference we also note
that $W_{T T}$ is not small
\be
W_{T T}^2 = 3 \sigma V_{\sigma\sigma} + \Or(\delta^2)
\label{Wtt}
\ee
at $\sigma=\sigma_0$. In the $a \sigma_0 \gg 1$ limit, a similar
relation also holds for KKLT: $W_{TT}^2 \approx  3 \sigma V_{\sigma\sigma}$.

\subsection{Chaotic inflation without a moduli sector}

In the simplest model of chaotic inflation the potential is just a
monomial, for example in quadratic chaotic inflation \cite{linde}
\be
V= \frac12 m^2 \vp^2 
\label{Vm}
\ee
with $\vp$ the canonically normalised real inflation field. Such a
model can be realized in a supersymmetric theory with a
superpotential~\cite{kawasaki}
\be
W= m \phi_1 \phi_2 \, ,
\ee
and defining the inflation field via $\vp= \sqrt{(|\phi_1|^2 + |\phi_2|^2)/2}$.  The equations
describing the perturbation spectrum are summarised in 
\ref{a:pert}, here we just mention the main results for quadratic chaotic inflation.  Inflation ends for
$\vp_e \simeq \sqrt{2}$.  Observable scales leave the horizon $N_* \sim 60$
e-folds before the end of inflation when $\vp_* \simeq 2 \sqrt{N_*}
\approx 15.5$. Here and in the following the subscript $*$ denotes the
corresponding quantity during observable inflation.  The spectral
index is $n_s -1 \simeq -2/N_* \approx 0.967$. Normalisation of  the
power spectrum to the observed values determines the mass scale 
$m \simeq 1.8 \times10^{13} \GeV$.  For future reference we also give the slow
roll parameters:
\be
\eta_*  = \epsilon_* \approx 8.4 \times 10^{-3} \, .
\label{srp}
\ee

More generically the potential will be some polynomial.

\subsubsection{Supergravity embedding}

Embedding chaotic inflation in supergravity gives corrections to the above SUSY model.  Explicitly:
\be
V_F= e^K\left(V_{\rm SUSY} + 2\Re [K^{\bar \imath j} K_{\bar \imath} \bar W
  \partial_j W] + [K^{\bar \imath j} K_{\bar \imath} K_j -3] |W|^2\right) \, .
\label{VF}
\ee

\paragraph{model 1}
The model is defined by
\be
K_{1} = - \frac12 (\phi_1 - \bar \phi_1)^2 + \phi_2 \bar \phi_2 \, ,
\qquad
W= m \phi_1 \phi_2 \, .
\label{model1}
\ee
The K\"ahler potential is invariant under a shift symmetry for the
inflaton field $\phi_1 \to \phi_1 +c$, which solves the
$\eta$-problem.  The shift symmetry is broken explicitly by the
superpotential, allowing for a small but finite inflaton mass.

We introduce the real canonically normalised fields $\phi_i = (\vp_i +
\rmi \alpha_i)/\sqrt{2}$\footnote{Arguably it is more natural to take $\phi_2 =
(\vp_2/\sqrt{2}) \exp(\rmi \alpha/\sqrt{2})$ for the canonically
normalised field, but since $\Im \phi_2 = 0$ in the minimum there is
no difference. For numerics the former definition is more useful as it does
not restrict $\vp_2$ to positive values.}.  It can be checked that the
potential is minimised for $\alpha_i = 0$ for parameter values of
interest.  The potential is then
\be 
V_F = \e^{\vp_2^2/2} \frac12 m^2
\(\vp_1^2 \left[1-\frac{\vp_2^2}{2} +\frac{\vp_2^4}{4} \right] + \vp_2^2
\)
\;\;\; \stackrel{\vp_2 \to 0}{\longrightarrow} \;\;\;
\frac12 m^2 \vp_1^2 \, .
\label{Vnomod}
\ee
For $\vp_2 = 0$, which is a stable minimum, all supergravity
corrections vanish and we retrieve quadratic chaotic inflation
\eref{Vm}.  Since the potential is steeper in the $\vp_2$-direction (no shift
symmetry), even with general initial values for both fields $\vp_2$
will be rapidly damped to zero, and inflation can commence.  A
potential problem with this model is that for small $\phi_2$ values, the
masses $m_{\vp_2}^2 = m_{\alpha_2}^2 =m^2$ are also light during inflation.
It was recently claimed that such a model may lead to large
non-Gaussianities during preheating~\cite{rajantie}.

\paragraph{model 2}

As a second explicit example we consider a model that is symmetric
under the interchange $\phi_1 \leftrightarrow \phi_2$, obtained by
also introducing a shift symmetry for $\phi_2$ in the K\"ahler
\be
K_{2} = - \sum_{i=1,2} \frac12 (\phi_i - \bar \phi_i)^2 \, .
\label{model2}
\ee
Decomposing again $\phi_i = (\vp_i + \rmi \alpha_i)/\sqrt{2}$ and setting
$\alpha_i =0$ in the minimum, the $F$-term potential is
\be 
V_F = \frac12 m^2 \left(\vp_1^2 +\vp_2^2 -\frac32 \vp_1^2 \vp_2^2\right) \, .
\ee
For $\vp_2 = 0$ the potential is that of quadratic chaotic inflation.
However $\vp_2$ is not a stable minimum for large $\vp_1$ values.
Although $\partial_2 V =0$ at $\vp_2=0$, the second derivative
$\partial_2^2 V = (m^2/4) (4 -6 \vp_1^2)$ turns negative for field
values $\vp_1 > 2/3$ as required for inflation.  Instead of rolling
towards the minimum $\vp_i=0$, the fields will run off to $\vp_i \to \infty$.  
The negative quartic term in $V_F$ is the cause of this
instability.  The quartic term comes from the last term in \eref{VF}.
Adding a no-scale modulus with $K = - 3 \ln(T + \bar T)$, which has
$K_T K^{T \bar T} K_{\bar T}=3$ (but for now without $T$ appearing in
the superpotential), the term cancels. The resulting potential is 
$V_F = (1/2) (T+\bar T)^{-3} m^2 (\vp_1^2 +\vp_2^2)$, perfect for
chaotic inflation provided $T$ is fixed somehow.  This model might
therefore work with a no-scale moduli sector, and we will look at it
in some detail in the next section.

\section{Chaotic inflation with a modulus sector}
\label{s:combine}

We now combine inflation with the modulus stabilisation sector~\cite{kallosh}.
To do so we simply add the respective K\"ahler and superpotentials 
\be
K =-3 \ln(T + \bar T) +  K_{a}(\phi_i,\bar \phi_i) \, , \qquad
W= \e^{\rmi \vartheta} \, W_{\rm mod}(T)+ m \phi_1 \phi_2 \, ,
\label{combine}
\ee
with $K_{a}$, $a=1,2$ the K\"ahler potential of the inflation model in the
absence of the moduli \eref{model1}, \eref{model2}.  $W_{\rm mod}(T)$ is
the non-perturbative superpotential that stabilises the volume
modulus; either a constant plus a single exponential as in KKLT
\eref{WKKLT}, or a modified racetrack potential with fine-tuned
parameters as in KL \eref{WKL}. We will take $W_{\rm mod}$ and $m$ to
be real. Any relative phase between the inflation and moduli sectors phase is
contained in $\vartheta$.

To assure that the modulus field does not run off to infinity during
inflation it has to be sufficiently heavy:
\be 
m_T^2 \gg H_*^2 = \frac13 V_* \sim 10^{-9}.
\ee
To get the second expression above we used the COBE normalisation for
the effective inflaton mass 
\be
\e^{K} m^2  \approx 6 \times 10^{-11}.
\label{cobem}
\ee

\subsection{Model 2}
\label{s:model2}

First consider Model 2 with a K\"ahler $\eref{model2}$ which is
invariant under shifts 
of both fields in the inflaton sector.   As mentioned in the introduction, this
model does not work.  However, it is interesting to see why.

Using \eref{model2}, \eref{combine} the potential 
can be written as $V = V_{\rm mod}(T) + V_{\rm inf}(T,\phi_i)+ V_{\rm
mix}(T,\phi_i)$, with $V_{\rm mod}(T)$  the moduli potential
after inflation when $\phi_i =0$, $V_{\rm inf}(T,\phi_i)$ the
inflaton potential in the limit $W_{\rm mod} \to 0$, and $V_{\rm
mix}(T,\phi_i)$ the remaining terms mixing the modulus and inflaton
sector. Introduce the real fields $\phi_i = (\vp_i + \rmi
\alpha_i)/\sqrt{2}$ and $T = \sigma + \rmi \gamma$.  In the regime of
interest the potential is minimised for $\alpha_i =0$.  Then
\bea 
V_{\rm mod}(T) &=& \frac{1}{6\sigma} 
\( |W_T|^2 - \frac{3\Re [W_T \bar W_{\rm mod}]}{\sigma} \) 
+ V_{\rm lift}(\sigma) \, , \nonumber\\
V_{\rm inf}(T,\vp_i)&=& 
\frac{1}{(2\sigma)^3} \frac{m^2}{2}  (\vp_1^2+\vp_2^2) \, , \nonumber\\
V_{\rm mix}(T,\vp_i) &=& 
 \frac{1}{(2\sigma)^3} \frac{m M(T)}{2} \vp_1 \vp_2 \, ,
\label{VII}
\eea
with
\be
M(T)= -4\sigma \Re [\e^{\rmi \vartheta} W_T] \, ,
\label{M}
\ee
$V_{\rm mix}$ is the correction to the inflaton potential due to the
presence of the moduli sector.  We will consider generic phases
$\vartheta$ for now.  Although fine-tuning the phase can make the
correction term $V_{\rm mix}$ arbitrarily small,  this is not enough
to save inflation.  As we will see shortly [see \eref{deltaV} below]
keeping the modulus $T$ as a dynamical field during inflation will
lead to instabilities in the inflaton potential, independently of
$\vartheta$.

In the KKLT scenario making the modulus heavy requires a large $W_T$,
and the correction term is 
large, ruining inflation.  For the model to work the correction term
should at least satisfy $M < m$ .  Then in the vacuum after
inflation when $V_{\rm mod} =0$, the inflaton mass eigenstates
$m_\pm^2 =m (m \pm M)$ are all positive definite.  In the
KKLT scenario $W_T T \sim W_0$, and $M$ can be made small decreasing
$W_0$. However, this also lowers the height of 
the  barrier ($3 m_{3/2}^2 \sim W_0^2$) separating metastable minimum
from the runaway minimum at infinity.  For generic phases there is no
parameter space where the moduli corrections are small $M < m$ yet the volume
modulus remains fixed $m_{3/2}^2 \gg H^2$.  This result
is independent of the specific form of the lifting term.

But the KL approach offers hope. A large stabilising modulus mass
requires $W_{TT}$ large while $M \propto T W_T \sim m_{3/2}$ can be
arbitrary small.  Indeed, after inflation when $\phi_i=0$ the
moduli and inflaton sector decouple in the mass matrix
\eref{massmatrix}, and by choosing suitable parameters, 
$m_T^2 \gg m_\vp^2 \gg m_{3/2}^2$ is possible.  We will give a
numerical example. Take $W_0 = w_0 +\delta$, see~\eref{w_0}, with
$A=2,\, B=2.4,\, a=0.2,\, b=0.25$.  Then $W_0=-0.079$, and the modulus
potential after inflation is minimised for $\sigma =8.1$ and 
$\gamma =0$.  The inflaton mass scale is set by the COBE normalisation
$m = 5 \times 10^{-4}$~\eref{cobem}.  The moduli masses
$m_\sigma^2 \approx m_\gamma^2 \approx 6 \times 10^{-7}$ are much 
larger than the nearly degenerate inflaton masses 
$m_{\rm inf}^2 \approx 6 \times 10^{-11}$.  The gravitino mass can be
made arbitrarily small  by taking $\delta \to 0$.  E.g.\ for 
$\delta = 10^{-10}$ we have $m_{3/2}^2 \sim 2 \times 10^{-14}$ and 
$E = 5 \times 10^{-17}$.  Then the moduli corrections in $V_{\rm mix}$
are very small as well $M/m \sim 10^{-10}$.

This looks all great, but the above masses are evaluated {\it after}
inflation, when $\vp_i =0$ and the mass matrix of the modulus and
inflaton sector decouple.  During inflation the correction term, the
mixing between the modulus field and the inflaton fields gives rise to
a tachyonic mode.  Consider for example $\vp_1 = 10$ and all other
fields at their instantaneous minima.  The moduli masses are practically the
same as in the vacuum, but the inflaton mass eigenstates (which have a
small admixture of the modulus) are now $m_\vp^2 = \{-9 \times 10^{-9},\, 5
\times 10^{-9},\, 9 \times 10^{-9},\, 6\times 10^{-11}\}$.  It is not
enough to have an eigenstate with $m_\vp^2 \ll H^2$, the other mass
eigenstates should be non-tachyonic as well.

\begin{figure}
\includegraphics[width=7cm]{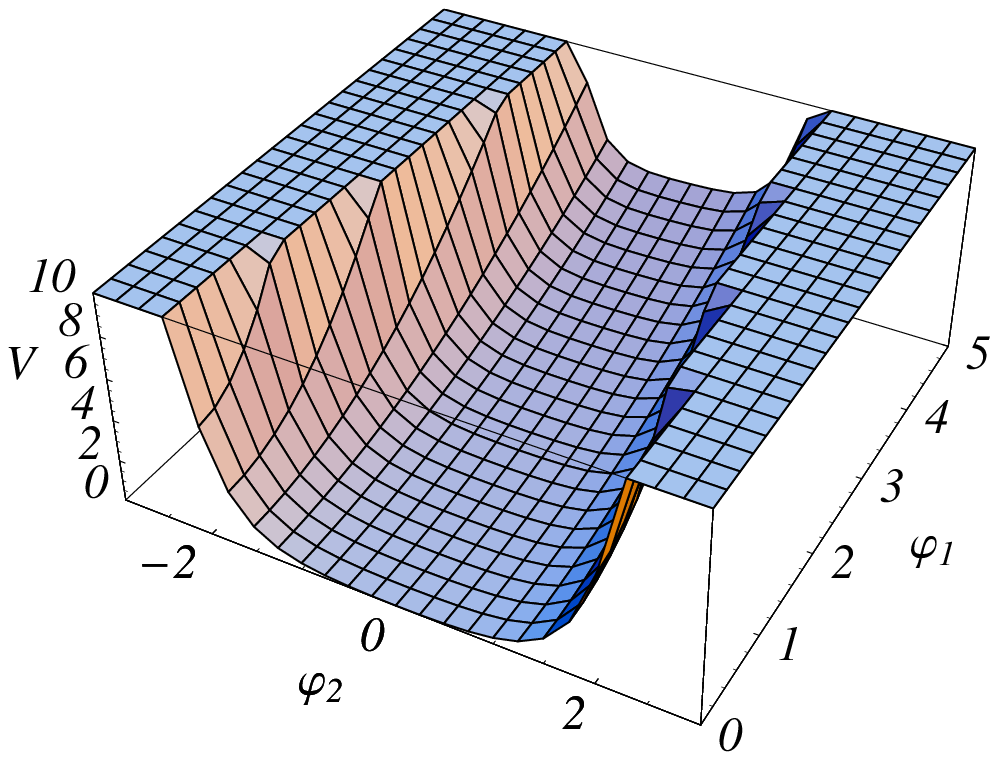} \hspace{0.2cm}
\includegraphics[width=7cm]{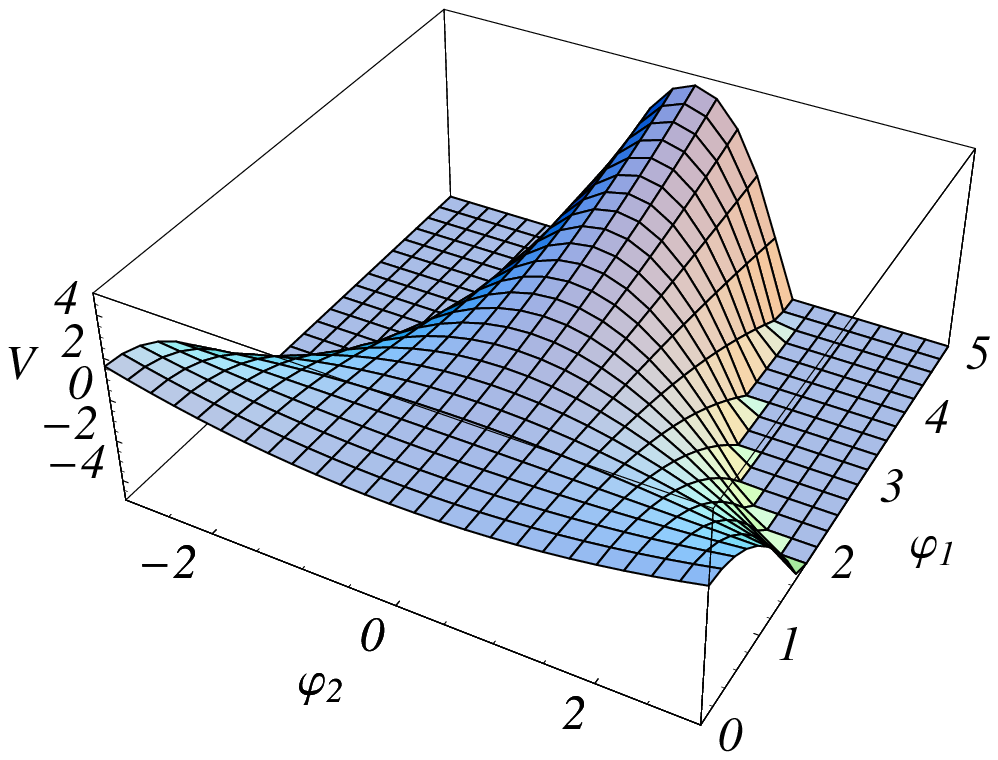}  \\
\caption{$V(\vp_1,\vp_2)$ (in rescaled units) with $\sigma$ at
  instantaneous minimum for  model 1 (left) and model 2 (right top).}
\label{F:potential}
\end{figure}

The origin of the tachyonic instability lies in the fact that for
non-zero $\vp_i$ the mass matrix has large off-diagonal terms 
$\propto (V_{\rm mix})_{\vp \sigma}\propto \vp W_{\sigma \sigma}$.
Although $W,\,W_\sigma\sim m_{3/2}$ are small in KL, $W_{\sigma \sigma}
\propto m_\sigma$ is not \eref{Wtt}.  There is related effect as well:
the $T$-field is displaced from the minimum.  Even though this
displacement is small, as a result the fine-tuned cancellation (tuning
at the level $\delta = 10^{-10}$ in our numerical example) in the KL
minimum no longer applies, and $W_T$ can be much larger than it is after
inflation.  As a result of all this the potential develops an
instability, no matter how small the gravitino mass in the
post-inflationary vacuum.  The effective potential as a function of
$\vp_1,\vp_2$ and with all other fields at their instantaneous minima
is shown in figure~\ref{F:potential}.

The appearance of the instability can be made explicit.  As remarked
above, $T$ is displaced from the post-inflationary minimum. 
Following~\cite{mfi} to estimate the effect of this we expand the
potential {\it during} inflation $T = \sigma_0 + \delta \sigma + \rmi \delta
\gamma$ with $T=\sigma_0$ the {\it post}-inflationary minimum. Then
\be \fl
V = V_{\rm inf} (\sigma_0,\vp_i) 
+ \sum_{x_A =\sigma,\gamma} \(\partial_A V_{\rm mix}(\sigma_0) \delta x_A + 
\frac12 \partial_A^2 V_{\rm mod}(\sigma_0) (\delta x_A)^2 \)
+ \Or\left((\delta x_A)^3, \frac{1}{\sigma_0},\delta\right)
\label{expand1}
\ee
where we have used that $V_{\rm mod}(\sigma_0) = V'_{\rm mod}(\sigma_0) =0$,
$\partial_\sigma \partial_\gamma V_{\rm mod} =0$ (the mass matrix is diagonal
in $x_A=\{\sigma,\gamma\}$), and we only kept the dominant terms.  Minimising
the potential with respect to $\delta x_A$ gives
\be
\delta x_A = - \frac{\partial_A V_{\rm mix}}{\partial_A^2V_{\rm mod}}
\qquad \Rightarrow \qquad
\delta V = \sum_{x_A =\sigma,\gamma}
\( -\frac12 \frac{({\partial_A V}_{\rm mix})^2}{\partial_A^2V_{\rm mod}}\) \, .
\label{deltaV}
\ee
Using the explicit form for $V_{\rm mix}$ \eref{VII} this evaluates
to a potential during inflation $V_{\rm eff} = V_{\rm inf} +\delta V$ with
\bea \fl
V_{\rm eff} & \approx & \frac12 \frac{m^2}{(2\sigma_0)^3}
\( \vp_1^2 + \vp_2^2 - 
\cos^2 (\vartheta) \frac{ W_{TT}^2}{(2\sigma_0) \partial_\sigma^2 V_{\rm mod}} \vp_1^2 \vp_2^2
- \sin^2(\vartheta)\frac{ W_{TT}^2}{(2\sigma_0) \partial_\gamma^2 V_{\rm mod}} \vp_1^2 \vp_2^2\)
\nonumber \\ \fl
&\approx & 
\frac12 \frac{m^2}{(2\sigma_0)^3}
\( \vp_1^2 + \vp_2^2 - \frac32 \vp_1^2 \vp_2^2\) \, ,
\label{Vinstab}
\eea
with $\vartheta$ the phase between the two sectors~\eref{combine}. To get the
bottom expression we used relation \eref{Wtt}, and the near degeneracy 
$\partial_\sigma^2 V_{\rm mod} = \partial_\gamma^2 V_{\rm mod} +
\Or(\delta^2)$ of the moduli masses.  
We see the appearance of a destabilising quartic term.  The mass
$m_2^2  \propto 1-(3/2) \vp_1^2$ becomes tachyonic during inflation
when $\vp_i = \Or(10)$,  and the potential develops an instability.
It is no use tuning the relative phases to set $V_{\rm mix} \to 0$;
the phase only determines whether it is the displacement of $\sigma$
or $\gamma$ that gives the largest correction, but its overall size is
phase-independent. For the above analysis to be valid, we need 
$\delta x_A \ll \sigma_0$, implying 
$m \vp_1 \vp_2 \ll m_\sigma \sigma_0^{3/2} $. This will hold if the
moduli stabilisation scale is much higher than the inflationary scale.

Ref.~\cite{brax} studied a similar model of SUGRA chaotic inflation
combined with a KKLT  moduli sector (they set $\vp_1 = \vp_2$),  and
found viable inflation for some parameters.  Their model avoids the
instabilities coming from the variation of $T$, although only for a
narrow range of the ratio $m/m_\sigma$. Our results
(\ref{expand1}-\ref{Vinstab}) do not apply to their model because in
their case the higher order terms,  both $\delta T/T$ and $1/T$,  are
not small.  However this comes at the cost of fine-tuning.
Furthermore, the value of $\sigma$ at the KKLT minimum does not
satisfy $\sigma_0 \gg 1$, and so the parameters used are on the
borderline of the validity of the model. Even so, this does give a
concrete example of how to evade~\eref{expand1} and its implications.  In
this paper we are more interested in chaotic inflation models that
work for general $m/m_\sigma$, without the need for tuning,  and will
describe such a setup in the next subsection.

To conclude, although the inflaton  and modulus sectors can be nearly
decoupled in the vacuum after inflation in the KL set-up, this is not true
during inflation.  The reason is that the (off-diagonal) corrections to the
mass matrix  $\propto (\partial_i W_{\rm inf}) W_{TT}$ are still large,
leading to a tachyonic direction in the potential.   To see this effect, it is
essential to treat the modulus $T$ as a dynamical field during inflation.
Even though the modulus displacement during inflation is small, it gives a
large correction to the inflaton potential which is crucial for a correct
analysis of the model.

\subsection{Model 1}
\label{s:model1}

In this section we discuss model 1 \eref{model1} combined with a moduli
sector.  As we will see, moduli corrections do not  destroy inflation, but
give small corrections which are potentially measurable.  Comparing with model
2 discussed in the previous section may give further insight into what
is needed for a successful inflation model in the presence of moduli.

Including a moduli sector the potential for model 1 \eref{model1},
\eref{combine} is
$V= V_{\rm mod} + V_{\rm mix} + V_{\rm inf}$ with
\bea \fl
V_{\rm mod} &=& \frac{\e^{\vp_2^2/2}}{6\sigma} 
\( |W_T|^2+ \frac{3 \Re (W_T
\bar W_{\rm mod})}{\sigma} \) + V_{\rm lift} \, ,
\nonumber \\ \fl
V_{\rm inf}  &=&\frac{\e^{\vp_2^2/2}}{(2\sigma)^3} 
\frac12 m^2 \Bigg[\vp_1^2 
\left(1+ \frac12 \vp_2^2\right)^2 +\vp_2^2  \Bigg] \, ,
\nonumber \\ \fl
V_{\rm mix} &=& \frac{\e^{\vp_2^2/2}}{(2\sigma)^3} \Bigg[
\frac12  m M \vp_1 \vp_2+  m
\Re[\e^{\rmi \vartheta} W_{\rm mod} ] \vp_1 \vp_2  \(1+ \frac{\vp_2^2}{2}\) 
+\frac{\vp_2^2}{2} |W_{\rm mod}|^2 \Bigg] \, ,
\eea
with 
$M$ given in \eref{M}.  We have set $\alpha_i =0$,
their values at the minimum during and after inflation in the
parameter regime of interest.  

The KKLT scenario does not work, due to the usual argument that it is
not possible to keep the moduli fixed during inflation while keeping
the soft corrections to inflation small.  Let us thus concentrate on
the fine-tuned KL model.    
One can calculate the effective potential, taking into account
the displacement of $T$ during inflation, analogous to
\eref{expand1}--\eref{Vinstab}.  The result is similar (note that the
effects of all extra terms appearing in $V_{\rm mix}$ in model 1
compared to model 2 give contributions $\propto W_{\rm mod}$ or
$\propto W'_{\rm mod}$ and are small in the KL set-up):
\be
V = V_{\rm inf} +\delta V
\approx
\frac12 \frac{m^2}{(2\sigma)^3} \e^{\vp_2^2/2}\left[ \vp_1^2\(1
+\frac12 \vp_2^2\)^2 + \vp_2^2 - 
\frac32  \vp_1^2 \vp_2^2\right]
\label{Vstab}
\ee
The difference between model 1 and 2 is that in model 1 the field
$\vp_2$ appears explicitly in the K\"ahler, and consequently $m_2^2$
receives additional stabilising contributions.  This is just enough to
keep the $\vp_2$-mass positive definite $m_2^2  \approx
m^2/(2\sigma_0)^3$; the $\vp_1$-dependent contribution to the mass
from $V_{\rm inf}$ and $\delta V$ cancels exactly.  A plot of
the potential as a function of $\vp_1,\vp_2$ and all other fields at
their instantaneous minimum is shown in figure~\ref{F:potential},
which confirms that the potential is stable during inflation; the
subdominant terms neglected in the analysis \eref{Vstab} do not affect
the stability.

Thus the moduli sector does not destabilise the inflationary
potential, and model 1 provides a viable model of SUGRA chaotic
inflation.   $V_{\rm eff}$ reduces to the quadratic chaotic
inflationary potential in the limit $\vp_2 \to 0$.  But $\vp_2 = 0$ is
not a minimum during inflation when $\vp_1 \neq 0$: $\partial_2
V|_{\vp_2 =0} = m \vp_1 (W_{\rm mod} -(T+\bar T) W_{T})/(T+ \bar T)^3
\neq 0$.   With KL moduli stabilisation the instantaneous minimum of
$\vp_2$ is close to zero.   As a result the potential is not exactly
quadratic but close to it.  To see whether these small deviations are
detectable,  we have integrated the equations of motion during
inflation numerically.  The relevant equations are given in
\ref{a:pert}.  The results are independent of the initial conditions
as long as $\vp_1$ is large enough for 60 e-folds of inflation;
$\vp_2$ (no shift symmetry) and $T$ are heavy and will soon settle in
their instantaneous minimum.

\begin{figure}
\vspace{1cm}
\includegraphics[width=7cm]{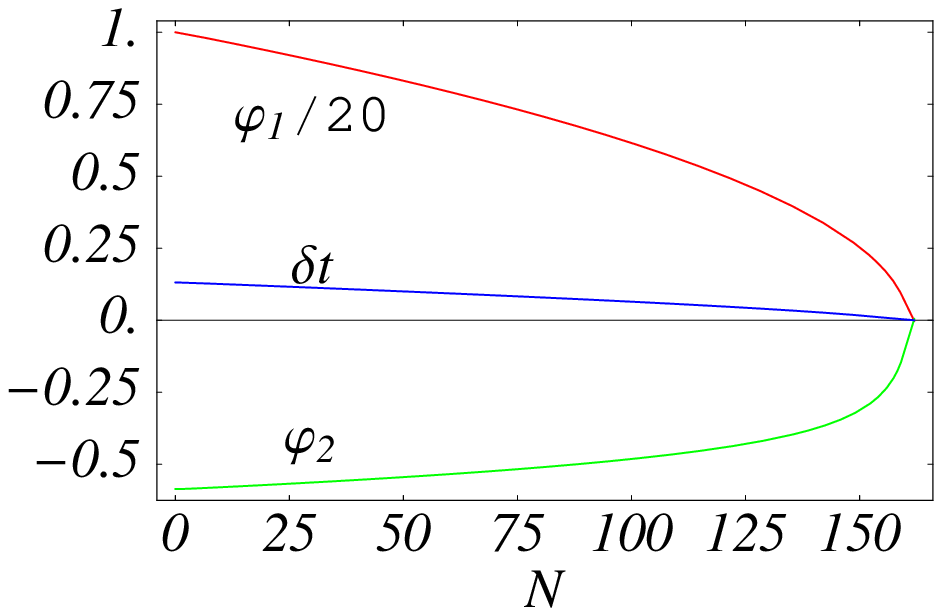} \hspace{0.2cm}
\includegraphics[width=7cm]{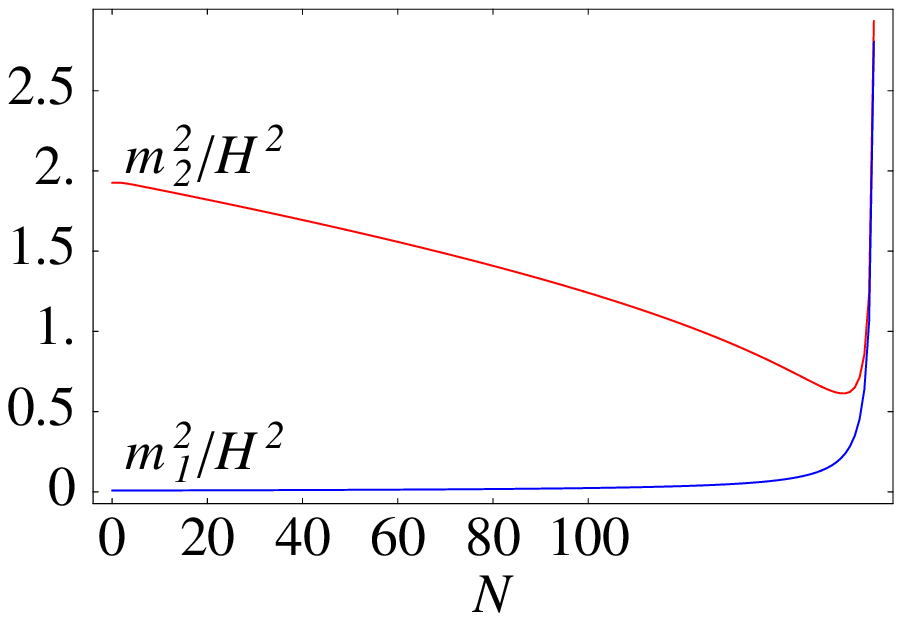} 
\caption{Results for model 1 with parameters $A=1,\,B=1.4,\,a=0.2,\,b=0.3,\, \delta = 10^{-8}$. Left plot shows field evolution of $\vp_1$ (rescaled), $\vp_2$ and $\delta T = T - T_0$ during inflation.  Right plot shows the inflaton mass eigenstates as function of time $N$.}
\label{F:evol}
\vspace{1cm}
\includegraphics[width=7cm]{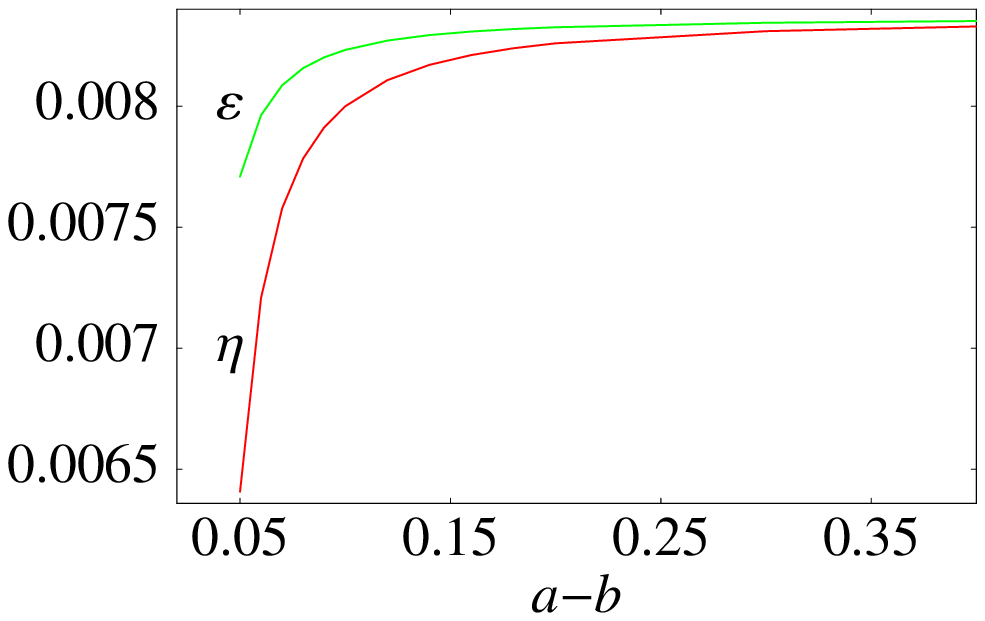} \hspace{0.2cm}
\includegraphics[width=7cm]{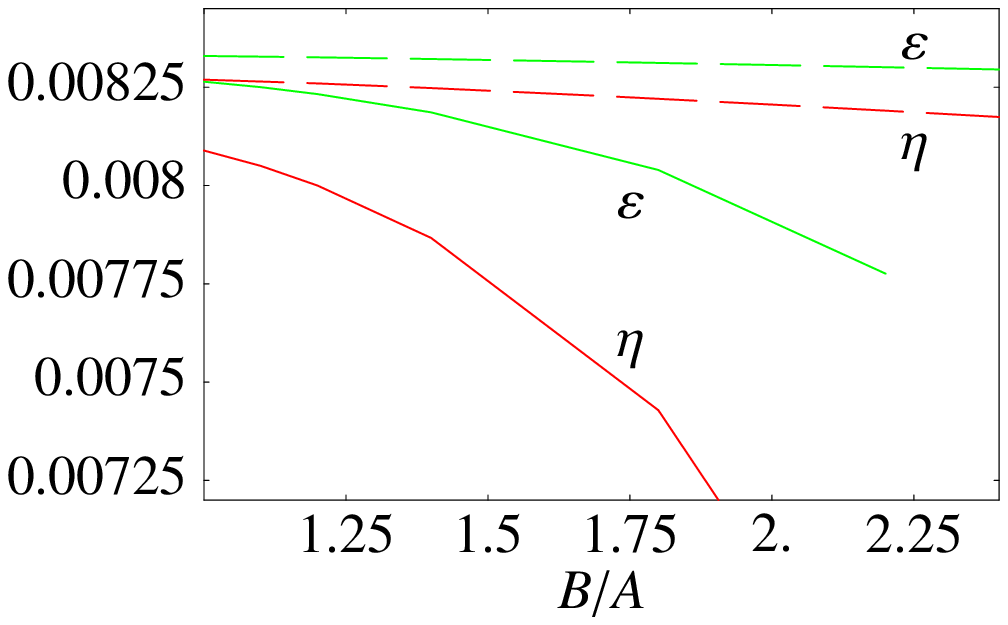} 
\caption{Slow-roll parameters.  Left plot shows $\eta,\epsilon$ as a function of $(b-a)$ 
for $A=1,\,B=1.4,\,a=0.2,\,b=0.3,\, \delta = 10^{-8}$; right plot shows $\eta,\epsilon$ as a function of 
  $B/A$ for the same parameters (solid lines) and for the $b=0.5$   (dashed lines).}
\label{F:slowroll}
\vspace{1cm}
\includegraphics[width=7cm]{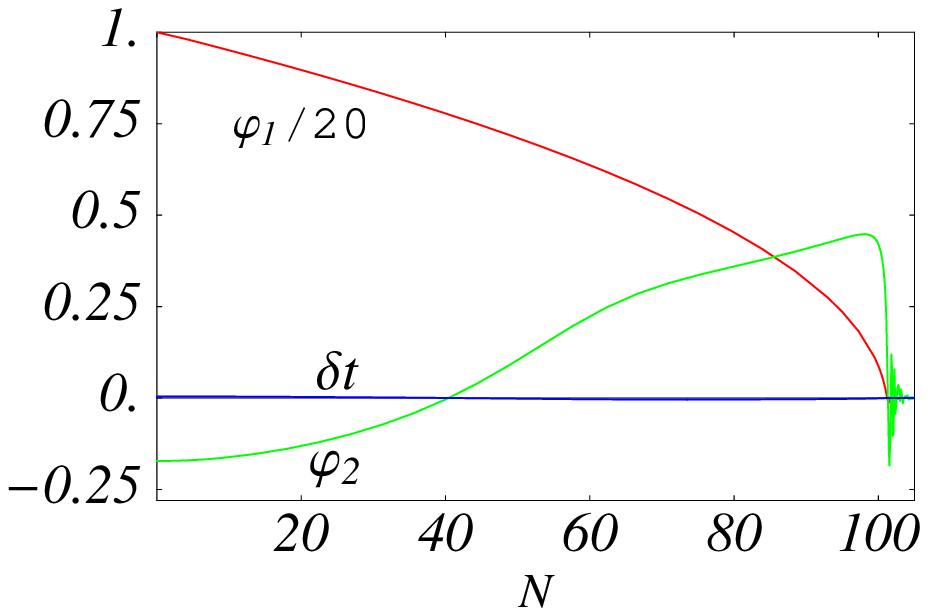}  \hspace{0.2cm}
\includegraphics[width=7cm]{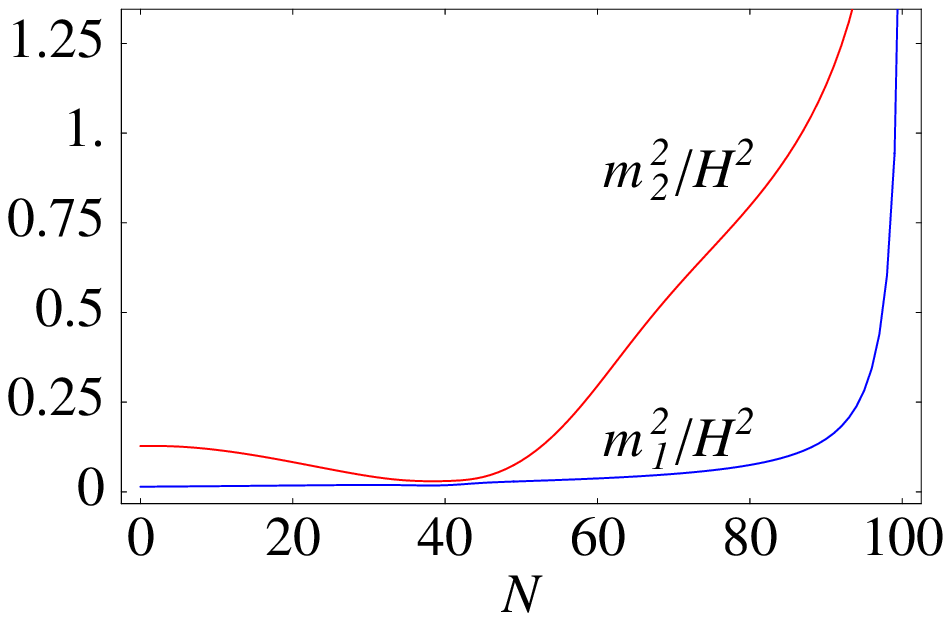}
\caption{Isocurvature modes.  Field values (left plot) and inflaton mass eigenstates (right plot) as a function of time $N$ for parameters $A=1,B=-1.2,a=0.5,b=0.66,\delta = 10^{-5}$.}
\label{F:iso}
\end{figure}

Consider a particular example with parameters
$A=1,\,B=1.4,\,a=0.2,\,b=0.3,\, \delta = 10^{-8}$. Then from
\eref{w_0} $W_0 \simeq -0.076$, and $E \simeq 2 \times 10^{-17}$ is
tuned to get a Minkowski vacuum after inflation.  The inflaton mass
$m=3.9 \times 10^{-4}$ is set by the COBE normalisation.  The field
evolution as a function of number of e-folds $N$ since the beginning
of inflation is shown in figure~\ref{F:evol}.  We started with 
$\vp_1 = 20$ and the other fields initially at their instantaneous minimum.
Inflation ends for $N \simeq 102$, observable scales leave the horizon
at $N \simeq 42$, when $(\vp_1)_* = 15.4$.

In the vacuum after inflation the inflaton and modulus masses are well
separated: $m_T^2 = 3.4 \times 10^{-5}$, and $m_{\rm inf}^2 = 4.2
\times 10^{-12}$.  The gravitino mass $m_{3/2}^2 = 3 \times 10^{-12}$
is small as a consequence of small $\delta$. The hierarchy is
preserved during inflation.  At COBE scales the modulus mass is $m_T^2
= 3.5 \times 10^{-5}$, the lowest mass eigenstate (predominantly
the shift symmetric $\vp_1$ with a small admixture of $\vp_2$ and $\sigma$)
is $m_{1}^2 = 3.8 \times 10^{-12}$, while the other inflaton field
is heavier $m_2^2 = 3.8 \times 10^{-9}$.  The two inflaton mass
eigenstates $m_i^2/H^2$ with $i=1,2$ are shown in figure~\ref{F:evol}.
The gravitino mass during inflation is $m_{3/2}^2 = 4.6 \times 10^{-7}$.

In all of parameter space (with one exception to be discussed
shortly) $m_2^2 > 0.1 -1 H^2$.  Although all fields evolve during
inflation, single field inflation is a good approximation.  We
calculated the slow-roll parameters projected along the inflaton
trajectory, and compared them with the usual slow-roll parameter in
terms of derivatives of the potential \cite{brax,gordon,tent1,tent2};
the difference is less than one percent.  See  \ref{a:pert} for the
relevant definitions.

The result for the perturbation spectrum are as follows.  The spectral
index in all of parameter space is $n_s = 0.967$, the same value as in
quadratic inflation.  The potential is not purely quadratic though.
In figure~\ref{F:slowroll} the slow-roll parameters are shown as a
function of $(b-a)$ (the results are fairly independent on absolute
scale $a$).  In the limit of large $(b-a)$ the slow-roll parameters
approach $\epsilon_* \simeq \eta_* \simeq 0.0083$ as for a purely
quadratic potential \eref{srp}, but they deviate for small $(b-a)$. If
tensor perturbations are observed in the future these deviations may
be measured, since
\be 
r =16 \epsilon, \qquad n_{_T} = -2\epsilon.
\ee
This breaks the degeneracy between a purely quadratic potential and
the current model with small $(b-a)$.  Figure~\ref{F:slowroll}
shows the slow-roll parameters as a function of $B/A$, for $a=0.2,\,
b=0.3$ and $a=0.2,\,b=0.5$ (lower and upper line).  The conclusion is that the
deviations from a purely quadratic potential can be large in the limit
$(b-a) \to 0$, $a \to 0$ and large $B/A$.  This is exactly the limit for which
$\sigma_0$ \eref{sigma0} is large.

The model works for $\delta \lesssim 10^{-4}$.  For larger $\delta$,
i.e.\  for a modulus sector with larger deviations from the Minkowski
SUSY minimum,  the mass eigenstates of the inflaton and moduli sector
can no longer be separated,  and the model is plagued by the same
problems as the KKLT set-up.  In the region $\delta \sim 10^{-4}-10^{-6}$
 there are parameters for which isocurvature fluctuations can
be large.  An example is shown in figure~\ref{F:iso}, for parameters
$A=1,\, B-1.2,\, a=0.5,\, b=0.66,\, \delta = 10^{-5}$.  The reason is
that for larger $\delta$ the field $\vp_2$ crosses the origin during
the inflationary evolution.  Around the origin the field is light
$m_2^2 \ll H^2$.  If this crossing happens around COBE scales, large
isocurvature fluctuations are produced.  This is the case for our
numerical example, where the origin crossing occurs around 60 e-folds
before the end of inflation when $N \approx 40$.  The evolution of
both the adiabatic and isocurvature perturbation is needed to
determine the spectrum; this is beyond the scope of this paper.

\section{KL moduli stabilisation and inflation}
\label{s:KL}

When can one successfully combine inflation with a fine-tuned KL-style
moduli sector (adding their respective superpotentials and only
coupling the two sectors gravitationally), and when not? The answer is
model dependent but the current discussion has gained some insight.
In this section we will expand on this some more. 

Consider a model with a K\"ahler potential
\be \fl
K = -3 \log \left[ T+\bar T - (T+ \bar T)^{\alpha} \frac{K_{a}(\phi_i,\bar
    \phi_j)}{3} \right] + K_{b}(\phi_i,\bar \phi_j) 
\equiv -3 \log X + K_{b} \, .
\label{Kall}
\ee
The source of instability comes from the terms
\be
V_{\rm mix} = \e^K [K^{T \bar T} K_T W_T \overline{W}_{\rm inf} +
 D_i W_{\rm inf} K^{i \bar T} \overline{W}_{\bar T} + {\rm c.c.} ] 
+ \cdots
\, ,
\label{Vmix_gen}
\ee
coupling the modulus and inflaton sector. Although $V_{\rm mix}$ is
small during inflation in the KL set-up\footnote{In fact $V_{\rm mix}$
can be tuned arbitrarily small by tuning the relative phase between
$W_T$ and $\overline{W}_{\rm inf}$.  However, as discussed in section
\ref{s:model2} the correction to the potential due to the dynamics of
the modulus field is independent of this phase, and thus cannot be
tuned.}, the off-diagonal corrections to the mass matrix $\propto
(V_{\rm mix})_{Ti} \propto (W_{\rm inf})_i W_{TT}$ are not \eref{Wtt}.
As a result, during inflation $T$ is slightly displaced from its
minimum $\delta T \propto \partial_T V_{\rm mix}$.  Although the
displacement is small, it disrupts the minute fine-tuning present in
the KL model, and as a result can lead to large corrections to the
inflaton potential.  This can be made explicit by Taylor expanding $T$
around its post-inflationary vacuum~\cite{mfi}.  The result is [see
\eref{expand1}--\eref{Vinstab}]  
$\delta V = - \sum (\partial_A V_{\rm mix})^2/ \partial_A^2V_{\rm mod}$ 
with $\delta x_A = \{\delta \sigma, \delta \gamma\}$. The minus sign
appears because $T$ will adjust to minimise the total potential.  The
effective inflationary potential is
\be
(V_{\rm inf})_{\rm eff} = V_{\rm inf} -  \sum_{x_A=\sigma,\gamma} 
\frac{(\partial_A V_{\rm mix})^2}{2\partial_{A}^2V_{\rm mod}} 
+ \Or \left(\delta, \frac1\sigma_0, \delta \sigma^3,\delta 
\gamma^3 \right) \, ,
\ee
with $V_{\rm inf}$ the potential in the limit that the moduli correction is
absent $W_{\rm mod} \to 0$.  The correction is potentially large, since
$\partial_A V_{\rm mix} \propto W_{TT}$, but model dependent.   The
superpotential could be a series of exponentials, or some polynomial in the
inflaton fields.  Here we have looked at polynomials, although we expect
similar results for both cases. The term $\delta V$ corrects the masses
of the inflaton sector fields.  For successful inflation the correction to the
inflaton mass needs to be sufficiently small so that $|\eta| \ll 1$.  But in
addition we have to make sure the masses of all other fields remain positive
definite during inflation, and the potential does not display an instability.
All mass corrections automatically vanish if $W_{\rm inf} =\partial_i W_{\rm
inf} =0 $ during inflation with $i$ running over all inflaton sector fields.
This is for example the case in $D$-term hybrid
inflation~\cite{D1,D2,mdi}.  But in all other cases the mass
corrections need to be checked, because as noted, they are large and
potentially destructive.

Consider first the case with $K_a=\partial_i K_a = \partial_i K_b=0$ during
inflation;  from~\eref{KTj} we see that the second term in 
$V_{\rm mix} \propto K^{i \bar T}$ vanishes, and the effective potential
becomes
\be
(V_{\rm inf})_{\rm eff} = 
 V_{\rm inf} 
-3 \frac{\e^{K_b}}{(2\sigma_0)^3} |W_{\rm inf}|^2 \, ,
\label{Veff_2}
\ee
where we have allowed for the variation of $T$ during inflation, and used
\eref{Wtt}. Take a superpotential $W_{\rm inf} \propto \lambda \phi$ linear
in the inflaton field.  The correction term in \eref{Veff_2} then alters
the inflaton mass.  Introducing a shift symmetry
for the inflaton $\phi$ to solve the $\eta$-problem, the moduli
correction can be calculated explicitly. It is too large: 
$\eta \simeq -3$.  An example is $F$-term hybrid inflation
\cite{mfi}. Thus KL with a linear inflaton superpotential that is
non-zero during inflation does not work .

Consider then  $W_{\rm inf} \sim \lambda \Pi_i \phi_i^{n_i}$ some polynomial
in inflaton sector fields. Now the correction term is a negative quartic or
higher order polynomial.  As before, inflation requires a sufficiently small
inflaton mass $|\eta| \ll 1$; this is automatic if $(W_{\rm inf})_{1} =0$ with
$\vp_1$ the inflaton field.  In addition all other ``spectator'' fields for
which $(W_{\rm inf})_i \neq 0$ should be non-tachyonic during inflation.  For
the chaotic inflation models discussed in this paper this is achieved if the
spectator field $\vp_2$ appears  in  $K_b$, as was the case for model 1.  
Note that the dominant mass correction to $\vp_2$ from $\delta V$ is the same for
all forms of the K\"ahler.  The reason that model 1 is stable and model 2 is
not, is simply that the inflaton potential $V_{\rm inf}$  in the former model
gives a larger stabilising contribution to the $\vp_2$-mass.

We will now consider a model with a more generic K\"ahler potential, for which
$K_a$ is non-zero \eref{Kall}, i.e.\ with the inflaton sector fields appearing
inside the log
\be
K_{\alpha} = -3\log\left[T+\bar T- \frac13 (T+\bar T)^\alpha 
\, \phi_2 \bar{\phi}_2\right]-\frac12 (\phi_1 -\bar{\phi}_1)^2 \, .
\ee
The most stable models are those for which $V_{\rm inf}$ gives the largest
mass to $\vp_2$.  Whether the shift-symmetric inflaton $\phi_1$ is inside or
outside the log does not affect the issue of stability --- for simplicity we
have put it outside the log in the K\"ahlers above.  However, how and where
$\phi_2$ appears is crucial.  If $\phi_2$ has a shift symmetry the potential
has an instability during inflation, $m_2^2 < 0$ for $\vp_1 \sim 1-10$,
whether $\phi_2$ appears outside the log in $K_b$ (as in model 2) or inside
the log in $K_a$.

As can be seen from the expressions in \ref{a:K}, the form
of the different parts of the potential ($V_{\rm mod}$, etc.) is rather
complicated. However we only require their leading order behaviour in
$\vp_2^2$. Furthermore, only the $W_T$ dependence part of $V_{\rm mix}$
will contribute significantly to $\delta V$~\eref{deltaV}. The
relevant terms are then
\bea 
V_{\rm mod} &=& \frac{1}{6\sigma}
\( |W_T|^2- \frac{3 \Re [W_T \bar W_{\rm mod}]}{\sigma} \) + \Or(\vp_2^2) \, ,
\nonumber \\ 
V_{\rm inf}  &=& \frac{m^2 \vp_1^2}{2(2\sigma)^{2+\alpha}}
+ \(1 +  \frac{2+4\alpha+\alpha^2}{6}\vp_1^2 \)
 \frac{m^2 \vp_2^2}{2(2\sigma)^3}+ \Or(\vp_2^4)
\, ,
\nonumber \\ 
V_{\rm mix} &=& \frac{1}{(2\sigma)^3} 
\frac{\alpha+2}{6}   m M \vp_1 \vp_2 + \Or(W_{\rm mod}, \vp_2^3)
\, .
\eea
Using the formula~\eref{deltaV}, we find 
$\delta V \approx -(2\sigma)^{-3} m^2\vp_1^2 \vp_2^2 (\alpha+2)^2/12$, giving
\be
m_2^2 = \frac{m^2}{(2\sigma)^3}\(1 -\frac{\vp_1^2}{3} \) \, ,
\ee
independently of $\alpha$.
What is rather surprising is that even $K_{\alpha}$ with $\alpha =1$
does not work. For small $\phi_2 \ll 1$ the log can be expanded  to
give $K_{\alpha} = K_{1} + \Or(|\phi|^4)$.
Since model 1 gives a viable model, one would expect $K_{\alpha}$ to
give similar results for $\alpha=1$.  But this is not the case.  The inflaton
potential $V_{\rm inf}$ differs for $K_1$ and $K_{\alpha}$, and thus
$\vp_2$ receives different stabilising mass contributions in each
case.   It is not enough to expand $K_{\alpha}$ first, and show that during
inflation $\phi_2$ is small to justify the expansion --- analysing the
full potential shows an instability.

We see that the placing $\phi_2$ inside or outside the log is crucial
to the success of inflation.  Model 1 with $\phi_2$ outside the log
gives a marginally stable model, where the $\vp_1$ dependent moduli
corrections just cancel.  As it turns out this is the most stable
model.  Placing $\phi_2$ inside the log, no matter what the modular
weight $\alpha$ is, gives a tachyonic mode.

\section{Conclusions}

In this paper we studied SUGRA chaotic inflation in the presence of
stabilised moduli fields.  To avoid the usual $\eta$-problem a shift
symmetry for the inflaton field is introduced.  But this is not
enough, as the moduli stabilisation sector gives rise to additional
contributions to $\eta$ and $\epsilon$ which are generically not
small.  The moduli sector breaks supersymmetry, and as a result the
inflaton fields get soft mass contributions of the order of the
gravitino mass.  These corrections need to be small for successful
inflation.  But in a generic moduli potential such as KKLT,  the
modulus mass is of the same order as the gravitino mass, and it is
impossible to keep the corrections to the inflaton small while making
sure the modulus remains fixed in its minimum during inflation.  KL
addressed this problem by constructing a fine-tuned moduli potential
with $m_{3/2}^2 \ll m_T^2$.  Indeed, calculating the potential in any
model in which inflation is combined with a KL moduli sector by adding
the respective superpotentials, the moduli corrections to inflation
appear small while at the same time the modulus is heavy.

All of the above assumes that  the modulus $T$ is fixed during inflation.
However, the modulus is a dynamical field, and this changes the
situation drastically.  Although during inflation the modulus is only slightly
displaced from its post-inflationary vacuum, this is enough to disrupt the
minute fine-tuning of the KL model.  The corrections to the effective inflaton
potential are generically large, and whether inflation works is a model
dependent question.   

Inflation combined with the KL moduli stabilisation scheme works well if the
derivative of the inflaton superpotential during inflation vanishes 
$(W_{\rm inf})_i =0$ with $i$ running over all inflaton sector fields. This is
for example the case for $D$-term hybrid inflation.  On the other hand if
$(W_{\rm inf})_i \neq 0$, there are large corrections to the masses of the
inflaton sector fields, which are missed if the modulus dynamics are not kept.
For models with $W_{\rm inf}$ a polynomial in the shift symmetric inflaton
field, these corrections are fatal.  If $W_{\rm inf}$ is some polynomial of
inflaton and ``spectator'' fields, the corrections to the $\eta$-parameter can
be harmlessly small if the spectator fields have a small VEV.  However, one
must also check that the masses of the spectator fields are positive definite
during inflation to avoid a run away behaviour.  For the chaotic inflation
models under consideration this requires the spectator field $\phi_2$ to have
a minimal K\"ahler (but note that this model is only ``just'' stable).  It is
not sufficient for $\phi_2$ to appear inside the modulus log with unit modular
weight, in which case upon a small field expansion  it will have a minimal
K\"ahler.  In fact, no matter what the modular weight, if $\phi_2$ is placed
inside the log [see \eref{Kall}] the spectator field becomes tachyonic during
inflation.

Our route to a successful inflation model in this paper was to take a 
specific choice of K\"ahler potential that minimises the impact of the
moduli corrections. We calculated the inflationary predictions for the
viable model 1, which has a minimal
kinetic term for the spectator field $\vp_2$ \eref{model1}, \eref{combine}.
Although the spectral index $n_s = 0.967$ is the same as for chaotic inflation
with a quadratic potential,  the values of the slow-roll parameters differ
from those of a purely quadratic potential.  The difference is largest for
those parameters that stabilise $T$ at large values.  The degeneracy between
the quadratic model and the model with moduli can be broken if tensor
perturbations are observed, as this allows us to extract the values of
$\eta$ and $\epsilon$ from the CMB data.  Hence, in the future, with
the launch of the Planck satellite, we may be able to observe the
presence of moduli fields in the sky.

Note that the problems arising from the variation of the modulus $T$
during inflation are not unique to chaotic inflation. Combining moduli
with $F$-term hybrid inflation was recently discussed in~\cite{dp},
where even a careful choice of K\"ahler could not save the
model. Instead, taking inspiration from~\cite{anaXW}, the moduli
problems were reduced by multiplying the superpotentials of the two
sectors, instead of adding them. It would be interesting to see if a
similar approach can help chaotic inflation models, although we will
leave this for future work.

\ack
SCD thanks the Netherlands
Organisation for Scientific Research (NWO) for financial support.

\appendix

\section{Perturbations}
\label{a:pert}

In this appendix we summarise the relevant equations for the perturbation
spectrum.

It is convenient to use the number of e-foldings $N =-\ln a$
(normalised so that $N = 0$ at the beginning of inflation) as a
measure of time. The scales measured by COBE and WMAP leave the
horizon $N_{\rm end} - N_* \approx 60$ e-folds before the end of inflation.
As before the subscript $*$ denotes the corresponding quantity at COBE
scales.  Slow-roll inflation ends when one of the slow-roll parameters
becomes greater than one.  In our numerical analysis we use $\epsilon =1$ to
determine the end of inflation.

To determine the inflationary trajectory, and the perturbation
spectrum, we integrate the equations of motion numerically, using
%
\bea
\frac{d \varphi_i}{d N} &=& \frac{1}{H} \dot \varphi_i(\pi_i), \nonumber \\
\frac{d \pi_i}{d N} &=& - 3\pi_i- \frac{1}{H} (V(\varphi_i)-\Lkin),
\label{eom}
\eea
with $\pi_i = \partial{\Lkin}/{\partial \dot \varphi_i}$. Dots indicate
derivatives with respect to $N$.

We can define the directional slow-roll parameter $\epsilon_\parallel$ as the
usual slow-roll parameter 
$\epsilon = (1/2) (V'/V)^2$ projected along the inflaton path \cite{brax,gordon,tent1,tent2}:
\be
\epsilon_\parallel = \frac{(\partial_N V)^2}{12 \Lkin V} \, .
\ee
We have checked that in all of parameter space (except for the case
with large isocurvature perturbations shown in figure~\ref{F:iso})
$|\epsilon - \epsilon_\parallel|/\epsilon < 10^{-2}$, and inflation is
effectively single-field with an adiabatic perturbation spectrum. 

The scalar power spectrum is then given by
\be
P = \frac{V}{150 \pi^2 \epsilon_\parallel}
\label{P}
\ee
evaluated 60 e-folds before the end of inflation. The COBE
normalisation imposes that $P \approx 4 \times 10^{-10}$.  A second
crucial observable is the spectral index of the inflaton fluctuations:
\be
n_s \approx 1 - \frac{d \ln P}{dN} .
\label{ns}
\ee
WMAP3 has measured $n_s =0.95 \pm 0.02$ for a negligible tensor
contribution to the perturbation spectrum~\cite{WMAP3}, and $n_s =0.98 \pm
0.02$ for non-zero $r$.  We checked that  using $n_s= 1 +2 \eta + 6 \epsilon$
instead to calculate the spectral index differs by less than a percent from
the spectral index \eref{ns}  of the adiabatic mode, confirming once again
that the usual single field equations apply.  The slow-roll parameter $\eta$
is defined as the minimum eigenvalue of the matrix
\be
N^a{}_b = \frac{g^{ac} (\partial_c \partial_b V
-\Gamma^e_{cb} \partial_e V)}{V} 
\ee
where the metric $g_{ab}$ is given by $\Lkin = (1/2) g_{ab}
\partial_\mu \varphi^a \partial^\mu \varphi^b$.

\section{General K\"ahler}
\label{a:K}

For a K\"ahler potential of the form 
\be
K = -\ln \left( T + \bar T - (T + \bar T)^\alpha
\frac{k(\phi_i, \bar \phi_i)}{3}\right) =
-3 \ln X
\ee
we find that $K_T = -[3+\alpha (T + \bar T)^{\alpha-1}k]/X$ and 
$K_i = (T + \bar T)^\alpha k_i/X$. The components of the inverse metric are
\be
K^{\bar T T} = \frac{X}{3{\cal C}}
\left(T + \bar T - (T + \bar T)^\alpha\frac{\tilde k}{3}\right)
\, , \qquad
K^{\bar T j} = (1-\alpha) \frac{X}{3{\cal C}} k^j \, ,
\label{KTj}
\ee
\be
K^{\bar \imath j} = \frac{X k^{\bar \imath j}}{(T+\bar T)^\alpha}
-\alpha\left(2-\alpha - (T + \bar T)^{\alpha-1}\frac{k}{3}\right) 
\frac{X}{3(T + \bar T){\cal C}} k^{\bar \imath} k^j \, ,
\ee
where $k^j =  \bar k_{\bar \imath} k^{\bar \imath j}$,
 $\tilde k = k - k^{\bar \imath j} \bar k_{\bar \imath} k_j$ and
\be 
{\cal C} = 1 -\frac{\alpha}{3}(T + \bar T)^{\alpha-1}\left[ k + 
\tilde k\left(2-\alpha - (T + \bar T)^{\alpha-1}\frac{k}{3}\right) \right]
\, .
\ee
For a minimal K\"ahler $\tilde k=0$, while for a shift
symmetry $\tilde k = -k$. We see that for all the models considered in
this paper, $\tilde k =0$ during inflation.


\section*{References}

\begin{thebibliography}{19}

\bibitem{linde}
  A.~D.~Linde,
  {\em Chaotic Inflation,}
  Phys.\ Lett.\  B {\bf 129} (1983) 177

\bibitem{kawasaki}
  M.~Kawasaki, M.~Yamaguchi and T.~Yanagida,
  {\em Natural chaotic inflation in supergravity,}
  Phys.\ Rev.\ Lett.\  {\bf 85} (2000) 3572
  [hep-ph/0004243]

\bibitem{kadota}
 K.~Kadota and M.~Yamaguchi,
  Phys.\ Rev.\  D {\bf 76} (2007) 103522
  [arXiv:0706.2676 [hep-ph]].

\bibitem{brax}
P.~Brax and J.~Martin,
  {\em Shift symmetry and inflation in supergravity,}
  Phys.\ Rev.\  D {\bf 72} (2005) 023518
  [hep-th/0504168]

\bibitem{kallosh}
  R.~Kallosh,
  {\em On Inflation in String Theory,}
  hep-th/0702059

\bibitem{kallosh2}
  R.~Kallosh and A.~Linde,
  {\em Testing String Theory with CMB,}
  JCAP {\bf 0704} (2007) 017
  [0704.0647 [hep-th]]

\bibitem{kss}
  R.~Kallosh, N.~Sivanandam and M.~Soroush,
  {\em Axion Inflation and Gravity Waves in String Theory,}
  0710.3429 [hep-th]

\bibitem{lyth}
  D.~H.~Lyth,
  {\em What would we learn by detecting a gravitational wave signal in
    the cosmic microwave background anisotropy?,}
  Phys.\ Rev.\ Lett.\  {\bf 78} (1997) 1861
  [hep-ph/9606387]

\bibitem{natural}
K.~Freese, J.~A.~Frieman and A.~V.~Olinto,
  {\em Natural inflation with pseudo - Nambu-Goldstone bosons,}
  Phys.\ Rev.\ Lett.\  {\bf 65} (1990) 3233

\bibitem{baumann}
  D.~Baumann and L.~McAllister,
  {\em A microscopic limit on gravitational waves from D-brane inflation,}
  Phys.\ Rev.\  D {\bf 75} (2007) 123508
  [hep-th/0610285]

\bibitem{bean}
  R.~Bean, S.~E.~Shandera, S.~H.~Henry Tye and J.~Xu,
  {\em Comparing Brane Inflation to WMAP,}
  JCAP {\bf 0705} (2007) 004
  [hep-th/0702107]


\bibitem{Nflation}
S.~Dimopoulos, S.~Kachru, J.~McGreevy and J.~G.~Wacker,
  {\em N-flation,}
  hep-th/0507205

\bibitem{grimm}
 T.~W.~Grimm,
  {\em Axion Inflation in Type II String Theory,}
  0710.3883 [hep-th]



\bibitem{assisted}
A.~R.~Liddle, A.~Mazumdar and F.~E.~Schunck,
  {\em Assisted inflation,}
  Phys.\ Rev.\  D {\bf 58} (1998) 061301
  [astro-ph/9804177]


\bibitem{copeland}
  E.~J.~Copeland, A.~R.~Liddle, D.~H.~Lyth, E.~D.~Stewart and D.~Wands,
  {\em False vacuum inflation with Einstein gravity,}
  Phys.\ Rev.\  D {\bf 49} (1994) 6410
  [astro-ph/9401011]

\bibitem{dine}
  M.~Dine, L.~Randall and S.~D.~Thomas,
  {\em Supersymmetry breaking in the early universe,}
  Phys.\ Rev.\ Lett.\  {\bf 75} (1995) 398
  [hep-ph/9503303]

\bibitem{gaillard}
  M.~K.~Gaillard, D.~H.~Lyth and H.~Murayama,
  {\em Inflation and flat directions in modular invariant superstring
  effective theories,}
  Phys.\ Rev.\  D {\bf 58} (1998) 123505
  [hep-th/9806157]

\bibitem{banks}
  T.~Banks, M.~Berkooz, S.~H.~Shenker, G.~W.~Moore and P.~J.~Steinhardt,
  {\em Modular Cosmology,}
  Phys.\ Rev.\  D {\bf 52} (1995) 3548
  [hep-th/9503114]

\bibitem{KL}
R.~Kallosh and A.~Linde,
  {\em Landscape, the scale of SUSY breaking, and inflation,}
  JHEP {\bf 0412} (2004) 004
  [hep-th/0411011]

\bibitem{KKLT}
S.~Kachru, R.~Kallosh, A.~Linde and S.~P.~Trivedi,
  {\em De Sitter vacua in string theory,}
  Phys.\ Rev.\  D {\bf 68}, 046005 (2003)
  [hep-th/0301240]

\bibitem{gkp}
  S.~B.~Giddings, S.~Kachru and J.~Polchinski,
  {\em Hierarchies from fluxes in string compactifications,}
  Phys.\ Rev.\  D {\bf 66} (2002) 106006
  [hep-th/0105097]

\bibitem{bkq}
 C.~P.~Burgess, R.~Kallosh and F.~Quevedo,
  {\em de Sitter string vacua from supersymmetric D-terms,}
  JHEP {\bf 0310} (2003) 056
  [hep-th/0309187]

\bibitem{ana}
 A.~Achucarro, B.~de Carlos, J.~A.~Casas and L.~Doplicher, {\em de Sitter
 vacua from uplifting D-terms in effective supergravities from
 realistic strings,} JHEP {\bf 0606} (2006) 014
 [hep-th/0601190]

\bibitem{GR}
  M.~Gomez-Reino and C.~A.~Scrucca,
  {\em Locally stable non-supersymmetric Minkowski vacua in supergravity,}
  JHEP {\bf 0605} (2006) 015
  [hep-th/0602246]

\bibitem{mfi}
P.~Brax, C.~van de Bruck, A.~C.~Davis and S.~C.~Davis,
  {\em Coupling hybrid inflation to moduli,}
  JCAP {\bf 0609} (2006) 012
  [hep-th/0606140]

\bibitem{O'Raifeartaigh}
  L.~O'Raifeartaigh,
  {\em Spontaneous Symmetry Breaking For Chiral Scalar Superfields,}
  Nucl.\ Phys.\  B {\bf 96} (1975) 331

\bibitem{rajantie}
  A.~Chambers and A.~Rajantie,
  {\em Lattice calculation of non-Gaussianity from preheating,}
  0710.4133 [astro-ph]

\bibitem{D1}
  E.~Halyo,
  {\em Hybrid inflation from supergravity D-terms,}
  Phys.\ Lett.\  B {\bf 387} (1996) 43
  [hep-ph/9606423]

\bibitem{D2}
  P.~Binetruy and G.~R.~Dvali,
  {\em D-term inflation,}
  Phys.\ Lett.\  B {\bf 388} (1996) 241
  [hep-ph/9606342]

\bibitem{mdi}
Ph.~Brax, C.~van de Bruck, A.~C.~Davis, S.~C.~Davis, 
R.~Jeannerot and M.~Postma,
  {\em Moduli corrections to D-term inflation,}
  JCAP {\bf 0701} (2007) 026
  [hep-th/0610195]

\bibitem{gordon}
  C.~Gordon, D.~Wands, B.~A.~Bassett and R.~Maartens,
  Phys.\ Rev.\  D {\bf 63} (2001) 023506
  [arXiv:astro-ph/0009131].

\bibitem{tent1}
  S.~Groot Nibbelink and B.~J.~W.~van Tent,
  Class.\ Quant.\ Grav.\  {\bf 19} (2002) 613
  [arXiv:hep-ph/0107272].

\bibitem{tent2}
  B.~J.~W.~van Tent,
  Class.\ Quant.\ Grav.\  {\bf 21} (2004) 349
  [arXiv:astro-ph/0307048].

\bibitem{dp}
S.~C.~Davis and M.~Postma,
  {\em Successfully combining SUGRA hybrid inflation and moduli stabilisation,}
  0801.2116 [hep-th]

\bibitem{anaXW}
 A.~Ach\'ucarro and K.~Sousa,
  {\em F-term uplifting and moduli stabilization consistent with Kahler
  invariance,}
  0712.3460 [hep-th]

\bibitem{WMAP3} 
D.~N.~Spergel {\it et al.}  [WMAP Collaboration],
  {\em Wilkinson Microwave Anisotropy Probe (WMAP) three year results:
  Implications for cosmology},
  Astrophys.\ J.\ Suppl.\  {\bf 170} (2007) 377
  [astro-ph/0603449]

\end{thebibliography}
\end{document}